\documentclass[aps,prd,twocolumn,superscriptaddress]{revtex4}

\usepackage{amsmath,amssymb,bm}
\usepackage{graphicx}
\usepackage{epstopdf}
\usepackage{amsfonts}
\usepackage{amssymb}
\usepackage{amsbsy}
\usepackage{amsmath}
\usepackage{latexsym}
\usepackage{sansmath}
\usepackage{subfigure}
\usepackage{lipsum}
\usepackage{float}
\usepackage{cancel}

\usepackage{bm}
\usepackage{xcolor}
 
\usepackage{comment}
\usepackage{csquotes}								
\usepackage{physics}
\usepackage{accents}

 \def\bea {\begin{eqnarray}}
\def\eea {\end{eqnarray}}
\def\nn {\nonumber}

\begin{document}
 
\title{Semiclassical geometrodynamics of homogeneous cosmology}

\author{Viqar Husain} \email{vhusain@unb.ca}
\affiliation{Department of Mathematics and Statistics, University of New Brunswick, Fredericton, Canada.}
\affiliation{Perimeter Institute for Theoretical Physics,
 31 Caroline Street North, Waterloo ON N2L 2Y5, Canada }

\author{Muhammad Muzammil} \email{muhammad.muzammil@unb.ca}
\affiliation{Department of Mathematics and Statistics, University of New Brunswick, Fredericton, Canada.}
 
\begin{abstract}
We study the classical-quantum (CQ) hybrid dynamics of homogeneous cosmology from a Hamiltonian perspective where the classical gravitational phase space variables and matter state evolve self-consistently with full backreaction. We compare numerically the classical and  CQ dynamics for isotropic and anisotropic models, including quantum scalar-field induced corrections to the Kasner exponents. Our results indicate that full backreaction effects leave traces at late times in cosmological evolution; in particular, the scalar energy density at late times provides a potential contribution to dark energy. We also show that the CQ equations admit exact static solutions for the isotropic, and the anisotropic Bianchi IX universes with the scalar field in a stationary state.  
 
\end{abstract}

\maketitle

\section{ Introduction} 
The road to quantum gravity (QG), albeit long and full of challenges, has provided some initial stepping stones. The well-studied area of quantum fields propagating on curved spacetime (QFCS) \cite{birrell1984quantum, parker2009quantum, fulling1989aspects} may be considered a first step; it applies where gravity is not dynamical. The next step is a theory where the important issue of backreaction of quantum matter on classical gravity is taken into account. This requires an idea for classical-quantum coupling. An attempt towards incorporating backreaction is the semiclassical Einstein equation
\begin{equation}
    G_{ab}+\Lambda g_{ab}= 8\pi G\ \langle \Psi | \hat{T}_{ab}(\hat \phi,\, g_{ab})|\Psi\rangle. 
    \label{sce}
\end{equation}
This is an equation for a semiclassical metric given a (Heisenberg) state $|\Psi\rangle$ of matter. It requires for its formulation a construction of the stress-energy tensor on an apriori undetermined metric. It is also (manifestly) non-linear in the state, unlike QFCS and QG---if it is presumed to be like usual quantum theory. Other issues with this proposal have been pointed out in the literature \cite{Isham:1995wr}. Despite these shortcomings, there have been many investigations of this equation (see e.g.  \cite{Page:1981aj,padmanabhan1989semiclassical,VachaspatiQBR,Ford:2005qz} and references therein). In addition, its applications to cosmology have been studied where the expectation value on the stress-energy tensor includes higher derivative corrections \cite{Simon:1991bm}. There is also recent work on this equation, again with higher derivative corrections where the metric is chosen to be static \cite{Juarez-Aubry:2020uim}. The standard approach used for studying (\ref{sce}) comes from noting that its right hand side, suitably renormalized, is in general a symmetric divergence free tensor $F_{ab}(g)$; this tensor inevitably has higher derivative curvature terms \cite{Ford:2005qz} computed to some order, with the accompanying issues that come with higher time derivatives. 

In contrast,  the work we report here uses a canonical approach starting from the Arnowitt-Deser-Misner (ADM) hamiltonian formulation for general relativity.  The idea is to formulate a ``semiclassical geometrodynamics" with effective constraints where the matter field is quantized and the gravitational phase space variables remain classical; the matter state evolves via a (functional) time dependent Schrodinger equation, and the gravitational phase space variables evolve via the effective constraints. As we show below, this system is self-consistent and provides a method to compute state and geometry evolution with full backreaction. The approach is similar to that  used recently in the so-called Friedmann-Schr\"{o}dinger equation \cite{Husain:2018fzg}, where a quantum state and scale factor evolve self-consistently from initial data via the Friedmann equation and the time-dependent Schr\"{o}dinger equation.   

To illustrate the approach and demonstrate its potential usefulness, we consider the case of scalar field coupled to homogeneous cosmology, both the isotropic and anisotropic cases. The general case of the latter with three scale factors moving in a potential has been the subject of much classical study beginning with the pioneering work of Belinskii, Lifshitz and Khalatnikov (BKL) \cite{BKL3}; further classical work on these models appears in \cite{ellis:1969,MacCallum:1971}; a survey of Hamiltonian cosmology, mainly in vacuum with a view to quantization appears in \cite{Ryan:1975jw}. To date there is no comprehensive study of the corresponding semiclassical system, at least  along the lines we present here.    

While our main interest in this paper is semiclassical cosmological dynamics with full backreaction, we also point out that there are static solutions of the classical-quantum equations. These are of interest in their own right and also due to past work, both in general relativity and in modified theories of gravity \cite{seahra2009einstein, boehmer2009stability, wu2011stability, Ellis:2002we,barrow2003stability, carneiro2009stability}; our approach may be viewed as a ``modified gravity theory," but not one with higher derivative terms. 

We begin in Sec. II with a numerical study of the isotropic case where we show the self-consistency of the classical-quantum (CQ) system, including preservation of the effective constraint under both state and geometric evolution, and conservation of probability;  we give dynamical solutions numerically, and static semiclassical solutions that resemble, but are different from the classical Einstein Static Universe; lastly, we compare classical and semiclassical evolution from comparable initial data. In Sec. III we generalize to the case of anisotropic  cosmology; we solve the semiclassical equations numerically to find quantum matter-induced corrections to the Kasner exponents and their classical sum rules; we also show that the Bianchi IX case admits static CQ solutions. We conclude in Sec. IV with a summary and discussion of further applications of this approach to semiclassical geometrodynamics.

\section{Classical-Quantum isotropic cosmology} 
The system we consider arises as a symmetry reduction of the Arnowitt-Deser-Misner (ADM) canonical action
\bea
S &=& \int dt d^3x \left\{ \pi^{ab}\dot{q_{ab}}+p_\phi \dot \phi-N\left(\mathcal{H}_G+\mathcal{H}_\phi\right) \right.\nn\\
\nn &&- \left. N^a\left(C^G_a+C^\phi_a\right)\right\}.
\eea
$(q_{ab},\, \pi^{ab})$ and $(\phi, \,p_\phi)$ are phase space variables for gravity and the scalar field respectively; $N$ is the shift, $N^a$ is the lapse; $C^G_a$ and $C^\phi_a$ are the diffeomorphism constraint functions;  ${\cal H}_G$ and ${\cal H}_\phi$ are the components of the Hamiltonian constraint:
\begin{equation}
\begin{aligned}
{\cal H}_G &=\frac{1}{\sqrt{q}}\left(\pi^{ab}\pi_{ab}-\frac{1}{2}\pi^2 \right)+\sqrt{q}\left(\Lambda - R^{(3)}\right)  \\
{\cal H}_\phi &=\frac{1}{2}\left(\frac{p_\phi^2}{\sqrt{q}}+
\sqrt{q}q^{ab}\partial_a\phi\partial_b \phi \right)+\sqrt{q}V(\phi)\\
C^G_a &=-2D_b\pi^b_a\\
C^\phi_a &=p_\phi \partial_a\phi;\\
\end{aligned}
\end{equation}
($\partial_a$ and $D_a$ are the spatial partial and covariant derivatives and we have set $8\pi G=1$ (to be reintroduced later)). 

The reduced canonical action for the 3-sphere FLRW model is given by the parametrization \cite{Ryan:1975jw} $q_{ab} =  a^2(t)\Sigma_{ab}$ and $ \pi^{ab}= \left(p(t)/6a(t)\right) \ \Sigma^{ab}\sqrt{{\rm det}\ \Sigma}$, where 
\bea
\Sigma_{ab}dx^adx^b = R^2\left(\sigma_1^2 + \sigma_2^2 + \sigma_3^2 \right)  
\eea
is the 3-sphere metric of curvature $\kappa = 1/R^2$ defined by the frame fields  $\sigma_1= \sin\psi\sin\theta d\phi +\cos\psi d\theta$, $\sigma_2 = \cos\psi\sin\theta d\phi -\sin\psi d\theta$, $\sigma_3 = \cos\theta d\phi +d\psi$, with $\theta\in[0,\pi]$, $\phi\in[0,2\pi]$, $\psi \in [0,4\pi]$. The ADM action becomes
\bea
S&=&V_0\int dt \left(p \dot a +p_\phi \dot \phi -N\left({\cal H}_G+{\cal H}_\phi \right)\right)\\
{\cal H}_G &=& -\frac{p^2}{24 a}+\Lambda a^3 -\kappa a, \\
{\cal H}_\phi &=&  \frac{p_\phi^2}{2a^3} + a^3 V(\phi)
\eea
where $V_0=\int \sqrt{{\rm det\ \Sigma}}= R^3\int \sigma_1\wedge \sigma_2\wedge\sigma_3$.  

Under the rescaling $R\rightarrow lR$, we have $\kappa \rightarrow \kappa/l^2$, $V_0\rightarrow l^3 V_0$, and the phase space variables scale as 
\bea 
a\rightarrow a/l,\  p\rightarrow p/l^2, \ \phi\rightarrow \phi, \  p_\phi\rightarrow p_\phi/l^3,  \eea
(the latter because $p_\phi$  is a density of weight one). Therefore it is useful to define the scale invariant variables
\bea
&&a\rightarrow a(V_0)^{1/3},\ \kappa\rightarrow \kappa(V_0)^{2/3},\nn\\ && p \rightarrow p(V_0)^{2/3}, \ p_\phi \rightarrow V_0p_\phi,   
\label{s-invar}
\eea
and the invariant volume $V\equiv V_0a^3$; the rescaled $a$ has dimension length and the rescaled $\kappa$ is dimensionless. The fundamental Poisson bracket becomes $\{a,p\}=1.$ We use these variables in the following.

Let us first see if there are classical static solutions. The equations of motion (with $8\pi G$ reintroduced)  are 
\bea
 {\cal H}&\equiv& {\cal H}_G + 8\pi G\  {\cal H}_\phi =0,\label{Hc}\\
\dot{a} &=& \{a,{\cal H} \} = -\frac{p}{12a}, \label{cadot}\\
 \dot{p}  &=& \{p,{\cal H} \}  
= -\frac{p^2}{24a^2} - 3\Lambda a^2 + \kappa \nn\\ 
&&+\ 24\pi G\left( \frac{p_\phi^2}{2a^4}
-  a^2 V(\phi)\right) \label{cpadot}\\
\dot{\phi} &=& \{\phi, {\cal H}\} =  \frac{p_\phi}{a^3},\\
\dot{p}_\phi &=& \{p_\phi, {\cal H}\}=-  a^3V'(\phi).\label{pphidot}
\eea
Let us consider the static ansatz
\bea
a&=&R={\rm constant} >0;\nn\\
\phi &=&\phi_0 = {\rm constant}; \quad p=p_\phi=0.\label{staticcons} 
\eea
Then eqn. (\ref{pphidot})  requires  $V'(\phi_0)=0$, i.e. $\phi_0$ must be an extremum of the potential, or the potential is a constant in a domain around $\phi_0$. With $v_0\equiv V(\phi_0)$  eqns. (\ref{Hc}) and (\ref{cpadot}) become 
\bea 
 \Lambda R^2 -\kappa +8\pi G R^2v_0   &=& 0,\nn\\
 -3\Lambda R^2 + \kappa-24 \pi GR^2 v_0 &=&0,
 \label{stateqns}
\eea
These have the solution
\bea
\Lambda =-8\pi G v_0,\quad \kappa =0,
\eea
that is, the cosmological constant is fixed by the potential at an extremum value $\phi=\phi_0$; this is of course flat space interpreted as an exact cancellation of the cosmological constant with $V(\phi_0)$.  However, if the potential is  quadratic, $V(\phi)=m^2\phi^2/2$,  eqn. (\ref{pphidot}) requires $ \phi_0=0$ due to the static ansatz (\ref{staticcons}), and  eqns. (\ref{stateqns}) have no non-trivial solutions.

To summarize the classical situation, we have seen that there are flat static solutions for non-quadratic potentials, but no solutions for the quadratic potential.  We will see that the latter is not the case for the classical-quantum equations we propose.

\subsection{Classical-Quantum equations: FRW}
We now describe the coupling of classical gravity with a quantized scalar field with quadratic potential. A canonical version for the model may be  defined in the Schr\"{o}dinger picture by starting with an effective Hamiltonian constraint where the scalar field is quantized, and deriving dynamics from such a constraint \cite{Husain:2018fzg};
the scalar field Hamiltonian operator is 
\bea
\hat{\cal H}_\phi = \frac{\hat{p}_\phi^2}{2a^3} + \frac{1}{2} a^3 m^2 \hat{\phi}^2.\label{Hop}
\eea
The proposed equations for the classical-quantum theory are  
\bea
{\cal H}_{\rm eff} &\equiv& -\frac{p^2}{24 a}+\Lambda a^3 -\kappa a + 8\pi G\ \langle \psi| \hat{\cal H}_\phi |\psi\rangle \nn\\
&=& 0,\label{heff}\\
i|\dot{\psi\rangle} &=& \hat{\cal H}_\phi|\psi\rangle,\label{tdse}\\
\dot{a} &=& \{a,{\cal H}_{\rm eff} \} = -\frac{p}{12a},\label{adot}\\
\dot{p}  &=& \{p,{\cal H}_{\rm eff} \} \nn\\
&=& -\frac{p^2}{24a^2} - 3\Lambda a^2 + \kappa  
 -  8\pi G \frac{\partial}{\partial a}  \langle \psi| \hat{\cal H}_\phi |\psi\rangle. \label{padot}
\eea
The last term of (\ref{padot}) may be expanded to give 
\bea
\frac{\partial}{\partial a} \langle \psi| \hat{\cal H}_\phi |\psi\rangle &=& 
\frac{1}{\dot{a}}\left(\dot{\langle \psi}| \hat{\cal H}_\phi |\psi\rangle + \langle \psi| \hat{\cal H}_\phi \dot{|\psi\rangle}\right) \nn\\
 && + \langle \psi| \frac{\partial}{\partial a} \hat{\cal H}_\phi |\psi\rangle \nn\\
&=& -\frac{3}{2} \left(\frac{\langle\hat{p}_\phi^2\rangle_\psi}{a^4}  - a^2m^2 \langle\hat{\phi}^2\rangle_\psi\right)\label{padotexpansion};
\eea
the first term is zero by (\ref{tdse}) and the last expression is a direct calculation from the scale factor dependence of $\hat{\cal{H}}_\phi$ in (\ref{Hop}). It is readily checked that the equations of  motion ensure conservation of the effective Hamiltonian constraint ${\cal H}_{\rm eff}$ (\ref{heff}):
\bea
\begin{aligned}
\frac{d}{dt} {\cal H}_{\rm eff} &= \frac{\partial \cal H_{\rm eff}}{\partial a}\dot a+\frac{\partial \cal H_{\rm eff}}{\partial p}\dot p\\
&+ 8\pi G\left(\langle \dot \psi | \hat{\cal H}_\phi | \psi \rangle+\langle \psi | \hat{\cal H}_\phi |\dot \psi \rangle\right)\\
&=-\biggr(-\frac{p^2}{24a^2} - 3\Lambda a^2 + \kappa \\
&-8 \pi G \frac{\partial}{\partial a}\langle \psi | \hat{\cal H}_\phi |\psi \rangle \biggr) \left(-\frac{p}{12a} \right)\\
&+\left(-\frac{p}{12a}\right)\biggr(-\frac{p^2}{24a^2} - 3\Lambda a^2 + \kappa \\
&-8 \pi G \frac{\partial}{\partial a}\langle \psi | \hat{\cal H}_\phi |\psi \rangle \biggr)
\\
&+8\pi G\left(i\langle \psi |\hat{\cal H}_\phi^2|\psi \rangle-i\langle \psi |\hat{\cal H}_\phi^2| \psi \rangle \right)\\
&=0.
\end{aligned}
\eea
It is also readily shown that probability is conserved. This ensures that the proposed system of equations is self-consistent. 

Initial data for this classical-quantum system at some $t=t_0$ is the set $\{a(t_0),p(t_0),|\psi\rangle(t_0)\}$  subject to the constraint ${\cal H}_{\rm eff} =0$; such data may be constructed by choosing $a(t_0)$ and  $|\psi\rangle(t_0)$, and solving ${\cal H}_{\rm eff} =0$ for $p(t_0)$. In this way the $p(t_0)$ is state dependent, and the free data is $\{ a(t_0), |\psi\rangle(t_0)\}$. In contrast, the constraint-free data for the classical system is $\{a(t_0),\phi(t_0),p_\phi(t_0) \}$, with $p$ determined by solving the hamiltonian constraint. Conservation of the semiclassical constraint (shown above) ensures, as in classical theory, that the evolved data continues to satisfy the constraint; we verify this explicitly in the numerical solutions presented below.

 \subsection{Static classical-quantum solutions}
 
Existence of static solutions of the classical-quantum equations may be checked by setting $a=R=$ constant $>0$ and $p=0$, and $|\psi\rangle =|n\rangle$, an eigenstate of $\hat{\cal H}_\phi$ (\ref{Hop}). For fixed $a=R$, 
\bea
\hat{\cal H}_\phi |n\rangle = E_n|n\rangle = m(n+1/2) |n\rangle.
\eea
This reduces  the evolution equations (\ref{heff}-\ref{padot}) to 
\bea
  \Lambda R^3 -R\kappa+ 8\pi Gm \left(n+1/2 \right)&=&0, \label{sc-static-1}\\
-3R^2\Lambda +\kappa &=&0 \label{sc-static-2}\\
|\psi\rangle(t) = e^{-iE_n(t-t_0)}|n\rangle,
\label{sc-static-3}
\eea
where the first equation comes from (\ref{heff}), the second from (\ref{padot}), and the third is the stationary state solution of the TDSE; eqn. (\ref{adot}) holds identically. Viewing the first two equations as conditions on $\Lambda$ and $\kappa$ gives the solutions
\bea
\kappa_n &=& 12\pi\ \frac{Gm}{R} \left(n+ \frac{1}{2} \right)= 12\pi G R^2\langle n|\hat{\rho}_\phi|n\rangle,\label{qc-static-kappa}\\
\Lambda_n &=& 4\pi\ \frac{Gm}{R^3} \left(n+ \frac{1}{2} \right) = 4\pi G \langle n|\hat{\rho}_\phi|n\rangle,
\label{qc-static-lambda}
\eea
where $\hat{\rho} = \hat{\cal H}_\phi/R^3$; (recall that $\kappa$ is dimensionless in the scale invariant variables (\ref{s-invar})). 

Thus, unlike the classical case with quadratic potential, static semiclassical solutions arise with   set values of both the cosmological constant and curvature; there are no static solutions with either $\kappa$ or $\Lambda$ being zero. The reason there is a semiclassical static solution and no classical one is that the expectation value $\langle n|\hat{\cal H}_\phi|n\rangle$ is independent of the scale factor (unlike the classical ${\cal H}_\phi$), therefore the Poisson bracket $\{ p, \langle n |\hat{\cal H}_\phi|n
\rangle\}=0$; this simplifies the $\dot{p}=0$ condition and permits a solution. Intuitively, an apparent reason is that there is a state of the massive scalar field in a $3-$sphere universe which does not spread due to the values of the cosmological constant and curvature.   

If the effective Hamiltonian constraint $(\ref{heff})$ is calculated in a coherent state $|\alpha\rangle$ of the scalar field, then $\langle\alpha| {\hat {\cal H}}_\phi|\alpha\rangle = m|\alpha|^2 + 1/2$, and again $\{ p, \langle \alpha|\hat{\cal H}_\phi|\alpha 
\rangle\}=0$; thus, the static semiclassical solution still arises, but now with (\ref{sc-static-1}-\ref{sc-static-3}) replaced by 
\bea
 &&\Lambda R^3 -R\kappa+ 8\pi Gm \left(|\alpha|^2+1/2 \right)=0, \\
&&-3R^2\Lambda +\kappa =0\\
&&|\alpha\rangle(t) = e^{-|\alpha|^2/2}
\sum_{n=0}^{\infty}e^{-iE_n(t-t_0)} \frac{\alpha^n}{\sqrt{n!}}|n\rangle.
\eea
The solution is the same as (\ref{qc-static-kappa}-\ref{qc-static-lambda}) with $n$ replaced by $|\alpha|^2$, and unlike the eigenstate case it allows any real values of $\kappa$ and $\Lambda$, but with both dependent on $|\alpha|$.

In summary, we see that there are no static solutions for coupling to a classical  massive scalar field, but there are for the classical-quantum case. For comparison, let us recall the static Friedmann equations $(a=R$) for perfect fluid equation of state $P=\sigma\rho$:
\bea
-\frac{4\pi G\rho}{3} (1+3\sigma) +\frac{\Lambda}{3}&=&0\\
\frac{8\pi G\rho}{3} +\frac{\Lambda}{3} -\frac{\kappa}{R^2}&=&0,
\eea
which lead to 
\bea
\Lambda &=& 4\pi G\rho(1+3\sigma);\\
\kappa &=& 4\pi G\rho R^2(1+\sigma).\label{statickappa}
\eea
It is therefore evident that classical static solutions arise for $\Lambda =0$ and $\kappa > 0$ if $\sigma=-1/3$; and $\Lambda > 0$ and $\kappa =0$  if $\sigma=-1$; the $\mathbb{R} \times S^3$ Einstein static Universe is the case $\Lambda > 0,\,\kappa > 0$ for any $\sigma\geq 0$. 

In contrast, the classical-quantum static solutions require $\Lambda > 0 $ and $\kappa > 0$ from eqns. (\ref{qc-static-kappa}-\ref{qc-static-lambda}); these exist for coherent states  and  any linear combination of energy eigenstates of the scalar field. Thus, these are like the Einstein static Universe but with $\Lambda$ and $\kappa$ determined by the scalar field state. 

In summary, for the massive scalar field, there are no classical static solutions, but there are classical-quantum static solutions.  

\subsection{Linear stability analysis} 

To check linear stability of the static solution ($a=R$, $p_a=0$, and $|\psi\rangle =|n\rangle)$, let
\begin{equation}
    \begin{aligned}
    a(t) &= R + \epsilon\, a_1(t)+ \mathcal{O}(\epsilon^2)\\
    p(t) &= 0 + \epsilon\, p_1(t)+ \mathcal{O}(\epsilon^2)\\
    \ket{\psi}(t) &= \ket{n} + \epsilon\, \ket{\chi}(t)+ \mathcal{O}(\epsilon^2),\\
    \end{aligned}
\end{equation}
with 
\bea 
|\chi\rangle(t) = \sum_k c_k(t) |k\rangle,
\eea
where $|k\rangle$ are eigenstates at the static point $a=R$; we insert these  
into the equations of motion (\ref{heff}-\ref{padot}). The effective constraint becomes
\bea
{\cal H}_{\rm eff} &=&  \Lambda R^3-\kappa R + 8\pi G m\left(n + 1/2 \right)\nn \\
&& +\ \epsilon\left[ 3\Lambda a_1R^2 -\kappa a_1 + 8\pi G\left(\langle n|\hat{\cal H}_\phi^{(1)}  |n\rangle \right.\right. \nn \\
&&  \left. \left. + \ \langle\chi|\hat{\cal H}_\phi |n\rangle
+ \langle n |\hat{\cal H}_\phi |\chi \rangle\right) \right],
\eea
where 
\bea
\hat{\cal H}_\phi^{(1)} = \frac{3}{2}\left(-\frac{a_1}{R^4}\ \hat{p}_\phi^2 + a_1R^2m^2 \hat{\phi}^2\right).
\eea
The first order constraint and evolution equations are 
 \begin{align}
    {\cal H}^{(1)}_{\rm eff}&\equiv  m(2n+1) {\rm Re}\{c_n\}=0,  \\
    \dot a_1&=-\frac{p_1}{12R}\label{a1dot}, \\
    \dot p_1&=-\left(6\Lambda R+\frac{9m}{R^2}\left(n+\frac{1}{2}\right)\right)a_1 -\frac{3m}{R} g(n),\label{p1dot}\\ 
    i \dot{c}_k&= E_k c_k + \frac{3ma_1}{2R}\bigg(\sqrt{n(n-1)}\ \delta_{k,n-2} \nn\\
    &\quad\quad +\sqrt{(n+1)(n+2)}\  \delta_{k,n+2} \bigg)\label{ckdot}, 
    \end{align}
where in (\ref{p1dot})
\bea 
g(n) &=& {\rm Re}\{c_{n+2}\}\sqrt{(n+1)(n+2)} \nn\\  
  && +\  {\rm Re}\{c_{n-2}\}\sqrt{n(n-1)}
\eea
\begin{equation*}
\begin{split}
g(n) &= \operatorname{Re}\left\{\sum_k c_k(t)\right\} 
\left(\sqrt{k(k-1)}\,\delta_{n,k-2} \right.\\
&\quad \left. + \sqrt{(k+1)(k+2)}\,\delta_{n,k+2} \right).
\end{split}
\end{equation*}
From (\ref{ckdot}) it is evident that only $c_{n-2}$ and $c_{n+2}$ evolve with a ``source,"  and all other perturbed state components ($ k\ne (n-2), (n+2) $) satisfy $c_k(t) = c_k(0)e^{-iE_k t},\ $. Furthermore, the perturbed constraint ${\cal H}^{(1)}_{\rm eff} =0$ is solved for all time if $c_n(0)=0$; i.e. the state perturbation excludes the static solution state $|n\rangle$, i.e. $ |\chi\rangle(t) = \sum_{k\ne n} c_k(t) |k\rangle$. Interestingly, this also ensures that probability is conserved to first order:
\bea 
\langle \psi|\psi\rangle &=& \langle n|n\rangle + \epsilon \left(\langle \chi|n\rangle + \langle n|\chi\rangle\right) + {\cal O}(\epsilon^2)\nn\\
&=& 1 + {\cal O}(\epsilon^2).\label{pcons}
\eea
For $n=0$ the above equations for linear perturbations reduce to a coupled set for $a_1,p_1$ and $c_2$; the equations for $c_1$ and $c_k,\ k>2$ have solutions $c_k = \exp{-iE_kt} c_k(0)$. The three eigenvalues for the former set depend on the values of the static solution parameters $m, R$ and $\Lambda$; a numerical check for a range of these parameters reveals that two of these are growing modes, and one is a decaying mode.  

\subsection{Comparison of exact classical-quantum and classical dynamics}

\begin{figure*} 
\begin{center}
   \includegraphics[width=\textwidth]{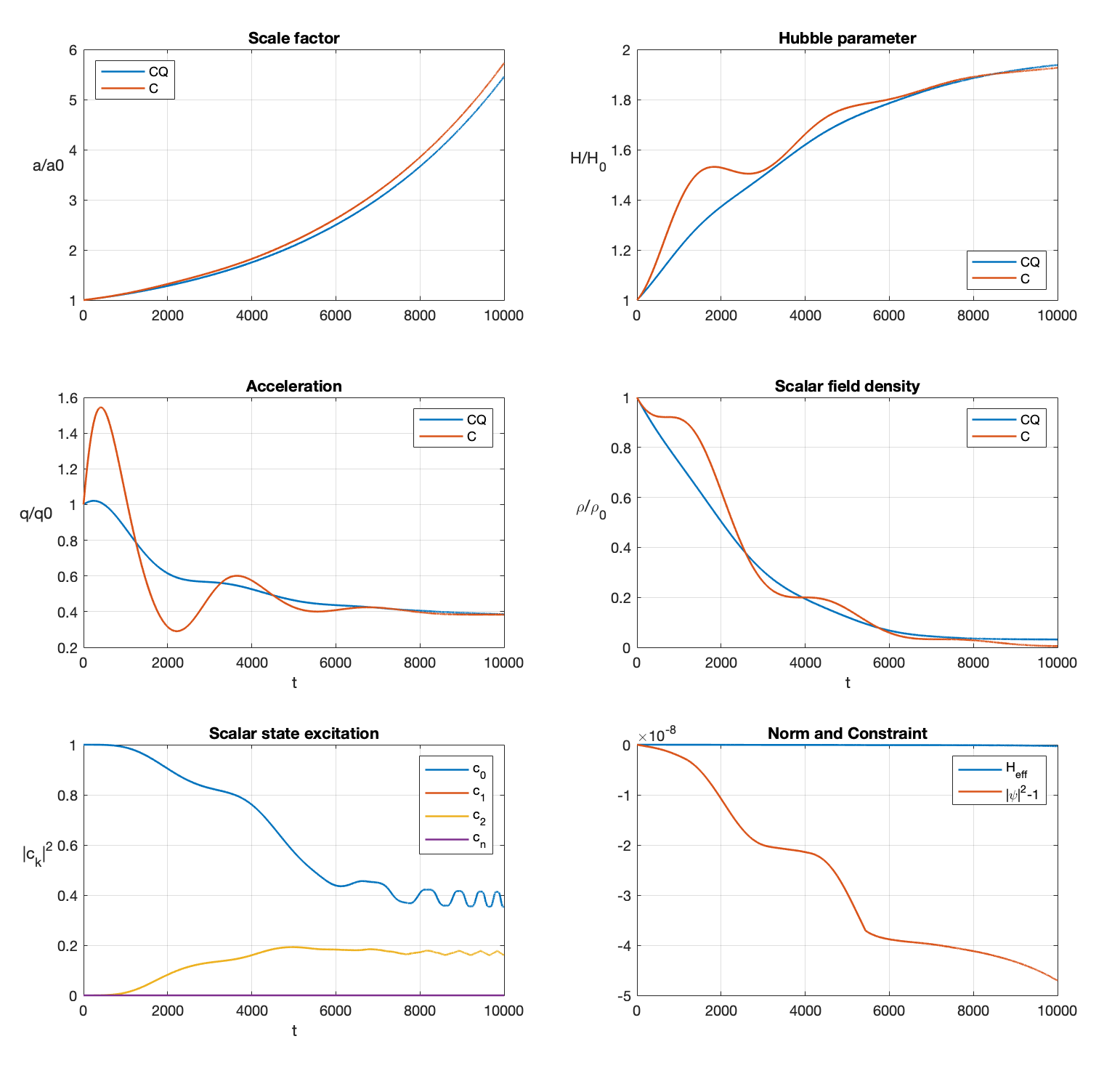}
  \caption{A typical comparative evolution of the classical-quantum (CQ) and classical (C) evolution with scalar field of mass $10^{-3}$ (Planck units) initially in its ground state $|0\rangle$. Initial data is a perturbation of a static semiclassical solution; the first four frames show the scale factor, Hubble parameter, acceleration, and density, relative to their initial values; the fifth frame shows the occupation probabilities  $|c_i|^2$ of higher scalar field states (with $c_n = c_{40}$); the last frame confirms conservation of probability and of the effective Hamiltonian constraint.}
  \label{fig1}
\end{center}
\end{figure*}

\begin{figure*} 
\begin{center}
   \includegraphics[width=\textwidth]{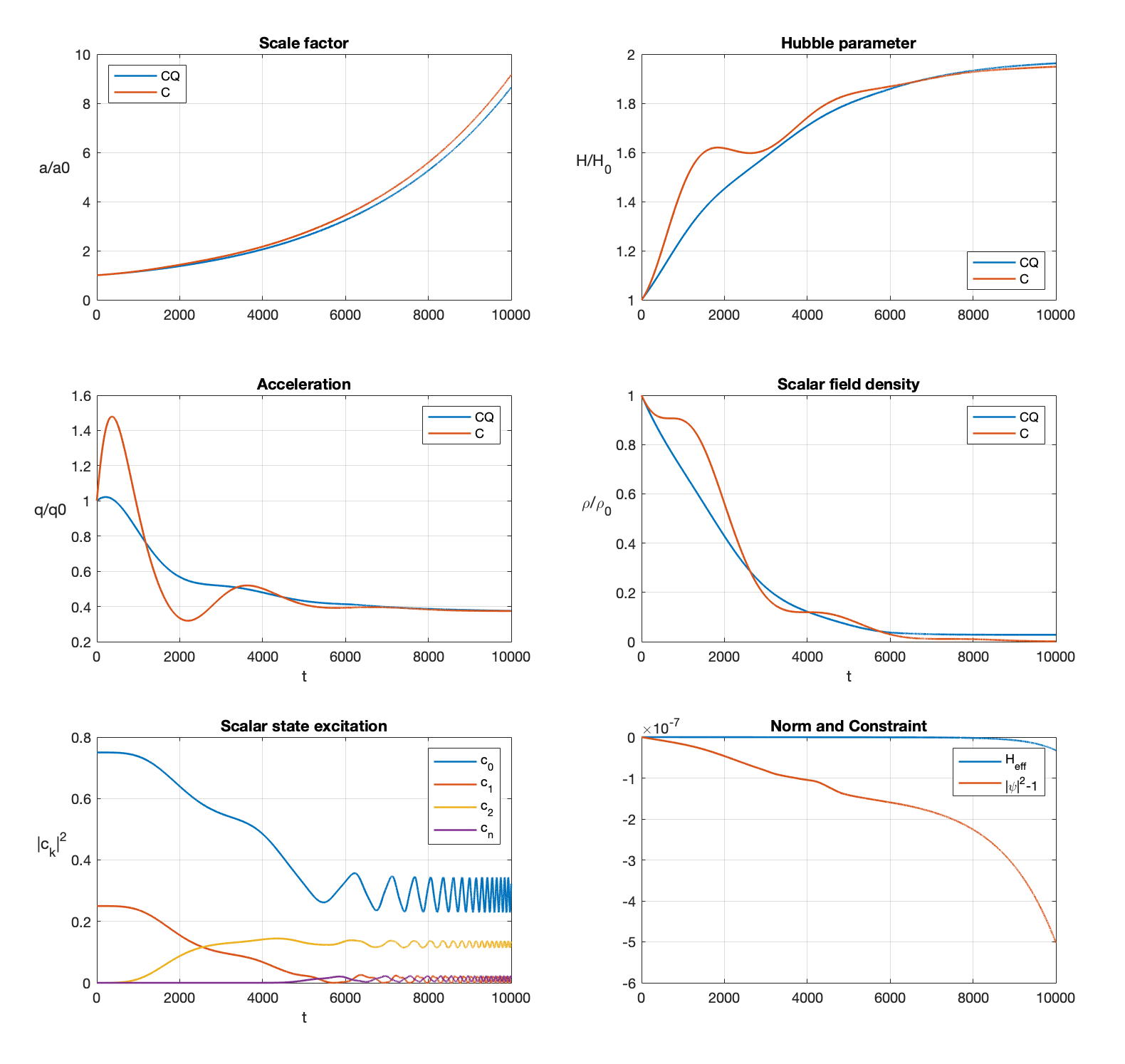}
  \caption{A typical comparative evolution of the classical-quantum (CQ) and classical (C) evolution for the same data as in Fig. \ref{fig1} but with scalar field initial state   $(\sqrt{3}/2)|0\rangle + (1/2)|1\rangle$. }
  \label{fig2}
\end{center}
\end{figure*}

In the last section we showed that there are exact static solutions of the classical-quantum system and discussed their linear stability. We now present  numerical solutions of the full classical-quantum equations (\ref{heff}-\ref{padot}) with initial data that is a small perturbation of the static classical-quantum universe; we compare these with the evolution of the classical-classical system (\ref{Hc}-\ref{pphidot}) with the same initial data. Specifically, we track the evolution of energy density, Hubble parameter, and acceleration in the two systems, while checking the conservation of the respective Hamiltonian constraints and the probability for the classical-quantum case.  

Suppose $\{a_s,|n\rangle;\Lambda_n,\kappa_n,m \}$ is a static solution of the classical-quantum equations. Then, the perturbed data we consider is the set
\bea
\{a_s + \Delta a, |n\rangle\}
\eea 
for the same parameter values $\{ \Lambda_n,\kappa_n,m \}$. 
Solving ${\cal H}_{\rm eff} = 0$ (\ref{heff}) with this perturbed data gives the initial value of $p$. From this we can compute  initial values for the observables of interest, the Hubble and acceleration parameters  $H=\dot{a}/a$ and $q= \ddot{a}/aH^2$, and energy density $\rho= \langle \hat{\cal{H}}_{\rm eff} \rangle_n/a^3$; the initial acceleration $q$ is proportional to $\dot{p}$ and is determined by the r.h.s. of eqn. (\ref{padot}). 

Initial data for the classical-classical equations is the set $\{a_s +\Delta a,\phi,p_\phi\}$, with $p$ determined by solving the Hamiltonian constraint. Matching the classical-quantum data requires the initial values of $\{\phi,p_\phi\}$ from the state $|n\rangle$. This is accomplished by setting 
\bea
\phi =\sqrt{\langle \hat{\phi}^2\rangle_n}; \quad
p_\phi =\sqrt{\langle \hat{p}^2_\phi\rangle_n},
\label{cdata}
\eea
a choice that matches the initial energy densities $\rho$ and accelerations $q$ for the CQ and C equations. The truncation of the scalar field Hilbert space used for numerical evolution is $n=40$, a number that does not affect probability conservation provided maximum evolution time is not too large and the chosen initial state is close to the ground state.

Fig. \ref{fig1} illustrates typical evolution for the classical-quantum and classical systems. Initial data is $a_0=a_s+5$ with $a_s=10$ with the scalar field initially its ground state $|n=0\rangle$; $m=10^{-3}$, $\Lambda=2.5\times 10^{-10}$ and $\kappa=7.5\times 10^{-6}$ (in Planck units) are fixed by the static solution for $a_s=10$. It is evident that the CQ and C evolutions differ, with the classical evolution of various quantities oscillating around the CQ one, and that at late times the CQ and C approach each other with small differences;  the fifth frame shows the excitation of higher states of the scalar field as the universe expands (with $c_n=c_{40}$, the truncation level of the oscillator basis used; the last frame demonstrates numerical conservation of probability and the effective hamiltonian constraint.

Fig. \ref{fig2} shows a similar evolution to that in Fig. \ref{fig1}, the only difference being the choice of initial state which is now $(\sqrt{3}/2) |0\rangle + (1/2) |1\rangle$. The comparative dynamics is nearly identical, except for the excitation probabilities of higher scalar field states. This occurs because the state evolution also connects all odd states in addition to even states as in Fig. 1. It is also evident that there is now also a small excitation of the last level in the truncation of the Hilbert space used for the numerical evolution. 

A variety of initial data that are perturbations of the semiclassical static solution yield qualitatively similar evolution. The agreement of the classical and classical-quantum results at late times provides justification for the classical-quantum  effective hamiltonian constraint. Lastly, instead of eigenstates of the scalar hamiltonian, we can also use coherent states, or arbitrary linear combinations thereof. The results are again qualitatively similar. The reason is that although the state evolves, $\langle \hat{{\cal H}}_\phi \rangle$ does not; the initial state only serves to fix the initial momentum $p$ conjugate to the scale factor via the effective constraint.     

\section{Classical-Quantum anisotropic cosmology}

In this section we extend the approach used above to the flat Kasner metrics
\bea
ds^2 = -N^2(t)dt^2 + \sum_{k=1}^3 \left(a_k(t)dx^k\right)^2  
\eea
With a scalar field the reduced ADM canonical action for this metric is  
\bea
S &=&\int dt\left(\sum_{k=1}^3 \pi_k\dot{a}_k + p_\phi\dot{\phi} - N\left( {\cal H}_K+{\cal H}_\phi\right)\right),\nn\\
 {\cal H}_K &=& \frac{1}{4v}\left[ \sum_k \lambda_k^2 - \frac{1}{2}\left(\sum_k \lambda_k\right)^2\right] + \Lambda v   \nn\\
  \cal{H_\phi} &=& \frac{p_\phi^2}{2v}+v V(\phi),
\eea
where $v= \prod_k a_k$ and $\lambda_k =a_k\pi_k$. 

For the vacuum case with $\Lambda=0$ and $N=1$, Hamilton's equations give  
\bea
\dot{\lambda}_i &=& \{ \lambda_i, {\cal H}_K\} =0, \label{ldot}\\
\dot{v} &=& \{v, {\cal H}_K\} = -\frac{1}{4}\sum_k \lambda_k \equiv \mu \label{vdot}\\
h_k &=& \frac{\dot{a}_k}{a_k}  = \frac{1}{2v} \left(\lambda_k + 2\mu\right) \label{hdot}.
 \eea
Thus, $\lambda_i$ are constants of the motion,  
\bea
v(t) = \mu t  + v_0
\eea
and the solution for the scale factors is 
\bea
a_k(t) = a_k(0) \left[ v(t) \right]^{p_k}, \quad p_k = 1+\frac{\lambda_k}{2\mu}.\label{Kexp}
\eea
It is readily verified from the definition of $\mu$ (\ref{vdot}) that the Kasner exponents $p_k$ satisfy 
\bea
\sum_k p_k = 1,\quad \sum_kp_k^2=1,
\eea
where the latter follows from the Hamiltonian constraint ${\cal H}_K=0$. These steps summarize the derivation of the well-known Kasner solution from the canonical equations. For non-zero scalar field and $\Lambda$, $\lambda_k$ are no longer constants of the motion as $\{\lambda_k, v\} =-v $. 

To see if there are classical static solutions with scalar field and non-zero $\Lambda$, we set $\phi=\phi_0=$ constant, $p_\phi=0=\pi_k$, and $a_k=a_{k0}=$ constants; in addition the r.h.s. of $\dot{p}_\phi$ and $\dot{\pi}_k = \{ \pi_k, {\cal H}_K  + {\cal H}_\phi\}$  must be set to zero. These conditions and the Hamiltonian constraint respectively give  
\bea
V'(\phi)|_{\phi_0}&=&0, \nn\\
\{\pi_k, v\} (\Lambda + V(\phi_0)) &=&0,\nn\\
v_0(\Lambda + V(\phi_0)) &=& 0.
\eea
Hence, there are non-trivial static solutions provided the potential has extrema $\phi_0\ne 0$ with tuned $\Lambda = -V(\phi_0)$.  Such solutions would have at least one linearly stable mode at extrema that are local minima, the double well being an obvious example; there is no non-trivial static solution for the purely quadratic potential.  

\begin{figure*} 
\begin{center}
   \includegraphics[width=\textwidth]{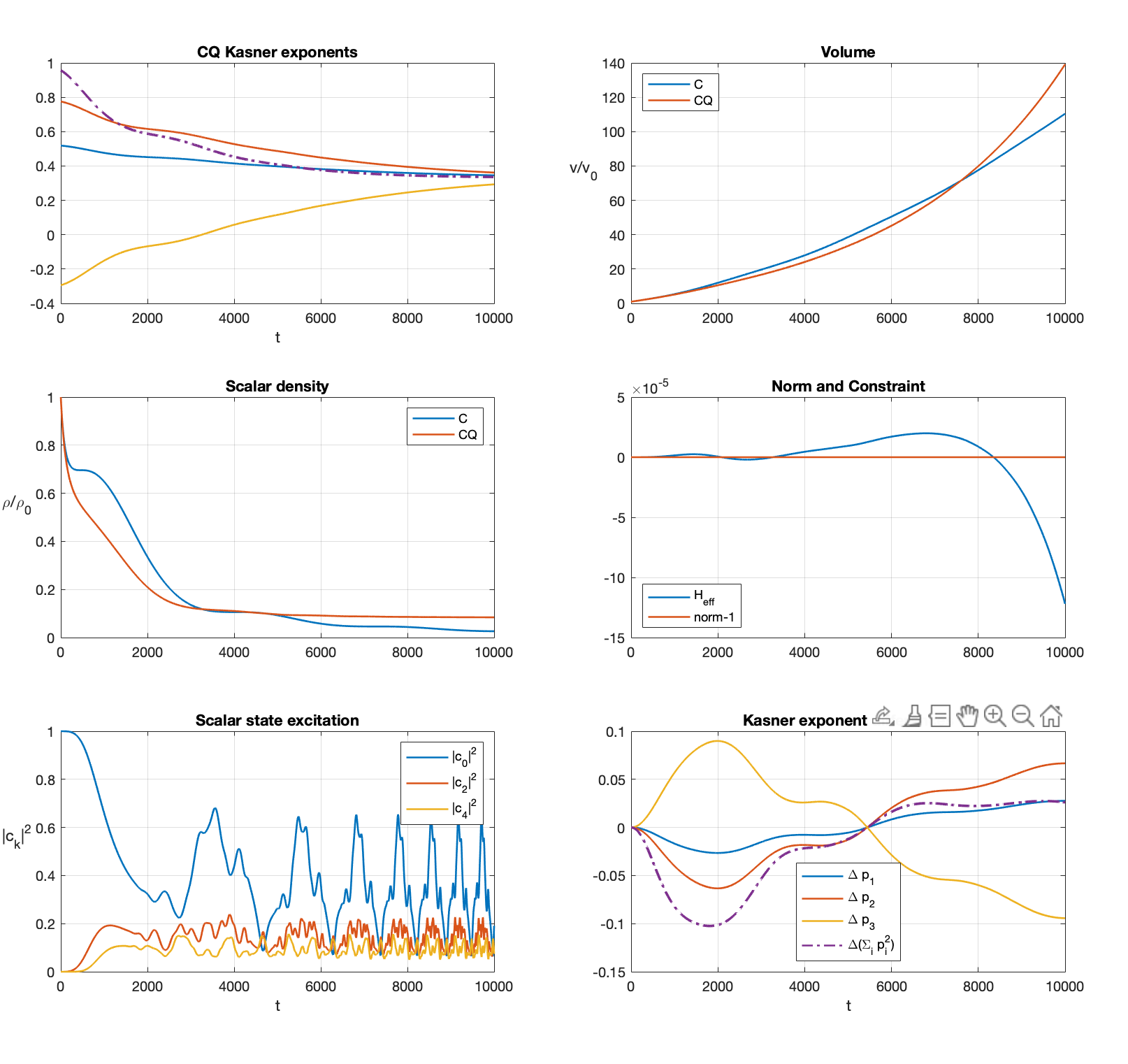}
  \caption{Numerical solution of the Kasner classical-quantum equations with the scalar field starting in the vacuum $|0\rangle$: the first frame shows the (dynamical) Kasner exponents, followed by comparison graphs  of volume and scalar field density; the forth frame shows conservation of the effective  constraint and probability; the last two frame show the excitation of the higher scalar states, and the difference between the classical and classical-quantum Kasner exponents.}
  \label{fig3}
\end{center}
\end{figure*}

\subsection{CQ Kasner}

We now consider a quantum scalar field coupled to Kasner spacetime. As for the FRW case discussed above we define the dynamics through an effective hamiltonian constraint
\bea 
{\cal H}_{\rm eff} \equiv {\cal H}_K + \langle \psi| \hat{\cal H}_\phi(v)|\psi\rangle.
\eea 
The proposed semiclassical geometrodynamics equations in the Schrodinger picture for the Kasner model are  
\bea
{\cal H}_{\rm eff}&=&0, \label{heffk}\\
i\dot{\ket{\psi}}&=&\hat{\cal H}_{\phi}\ket{\psi}(v),\label{psikdot}\\
\dot{a}_k &=& \{a_k, {\cal H}_K\},\label{akdot}\\
\dot \pi_k &=&\left\{\pi_k, {\cal H}_K + \langle \psi| \hat{\cal H}_\phi(v)|\psi\rangle \right\} \label{pikdot}.
\eea
The scalar field Hamiltonian operator is 
\bea
\hat{{\cal H}}_\phi = \frac{\hat{p}_\phi}{2v} +vV(\hat{\phi})
\eea
Evolution of $\pi_k$ is explicitly state dependent through the factor $v$ in $\hat{\cal H}_\phi$, and that of the scale scale factors $a_k$ is implicitly through $\pi_k$. As for the FRW case above,  state evolution is in the Schrodinger picture in a truncated fixed oscillator basis $\displaystyle |\psi\rangle=\sum_k^N c_k(t) |k\rangle$. 

\begin{figure*} 
\begin{center}
   \includegraphics[width=\textwidth]{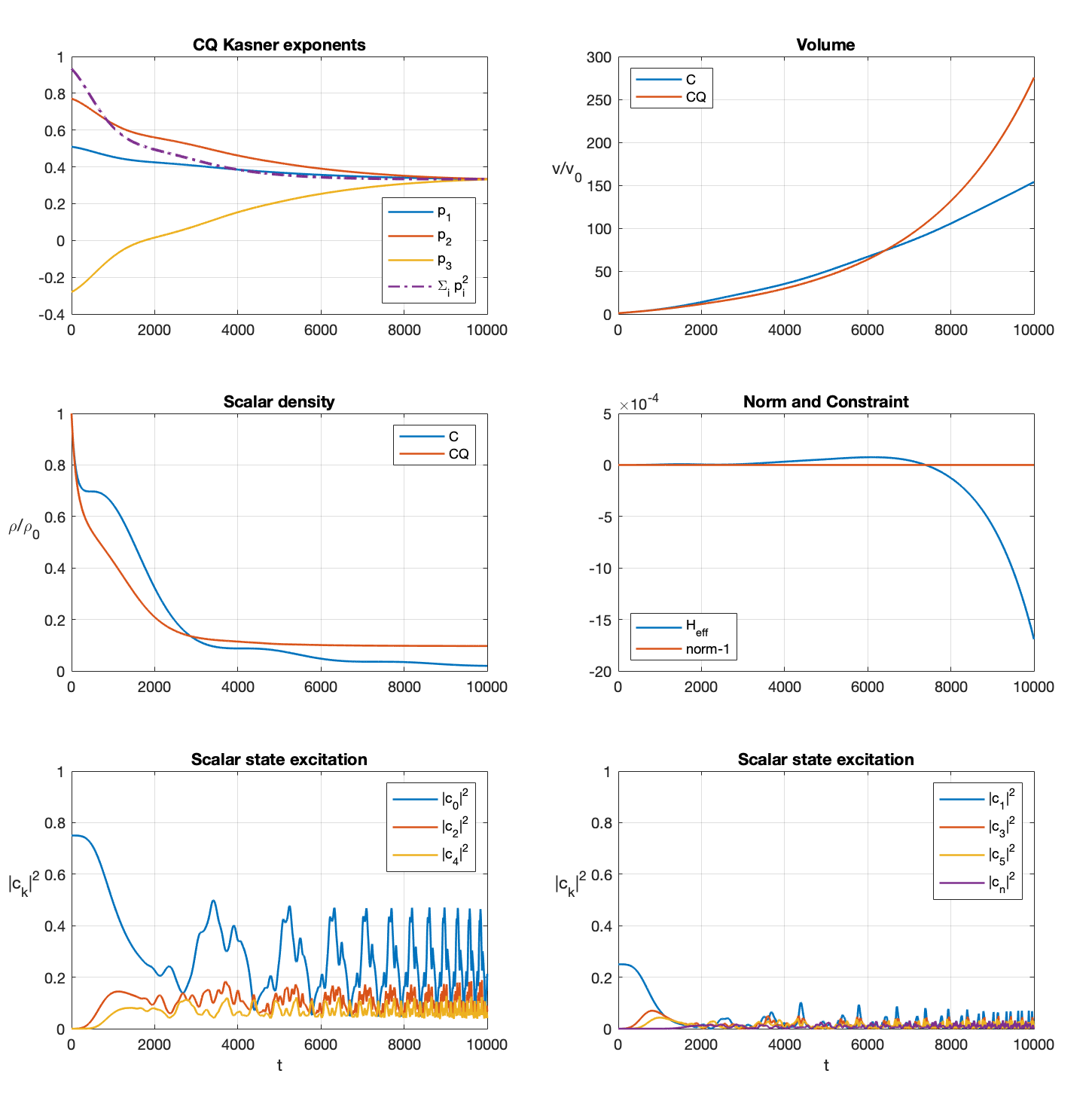}
  \caption{Numerical solution of the Kasner classical-quantum equations with the scalar field starting in the state $(\sqrt{3}/2)|0\rangle + (1/2)|1\rangle$: the first frame shows the (dynamical) Kasner exponents for the CQ case, followed by comparison graphs  of volume and scalar field density; the last two frames show the excitation of the higher scalar states;  $c_n=c_{50}$ is the truncation level of the scalar field basis.}
  \label{fig4}
\end{center}
\end{figure*}

\begin{figure} 
\begin{center}
   \includegraphics[width=\columnwidth]{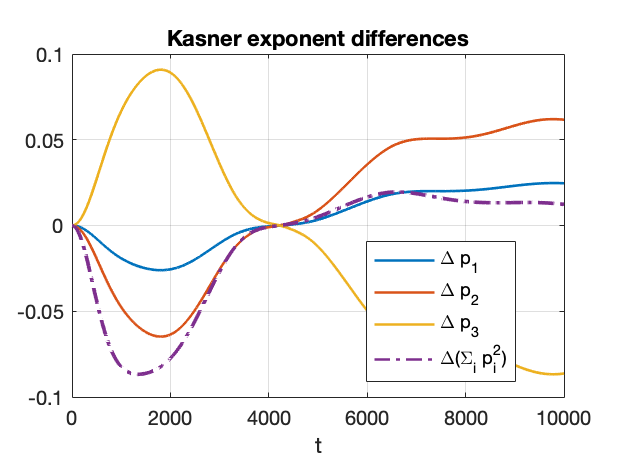}
  \caption{The difference between the CQ and C dynamical Kasner exponents for the data used for Fig. \ref{fig4}. } 
  \label{fig5}
\end{center}
\end{figure}

It is readily verified that ${\cal H}_{\rm eff}$ is conserved using the evolution equations:

\begin{widetext}
\[
\frac{d}{dt}{\cal H}_{\rm eff}= 
\sum_{k=1}^3\left[ \frac{\partial {\cal H}_{\rm eff}}{\partial a_k}\dot{a}_k+\frac{\partial {\cal H}_{\rm eff} }{\partial \pi_k}\dot{\pi}_k \right]  +\ \langle \dot\psi | \hat{\cal H}_\phi |\psi\rangle 
+  \langle \psi | \hat{\cal H}_\phi |\dot \psi\rangle=0.
\]
\end{widetext}

\subsection{Numerical solution}

Let us first note that there are no static solutions of the above equations: setting the scale factors to constants and their conjugate momenta to zero, the effective Hamiltonian constraint in an eigenstate of ${\cal H}_\phi$ and the $\dot{\pi}_k$ equations are 
\bea
\Lambda v_0 + m(n+1/2) &=& 0,\nn\\
\{\pi_k, {\cal H}_K\}|_{\pi_k=0} &=& v_0\Lambda/a_k;
\eea
the first gives $\Lambda = -m(n+1/2)/v_0\ne 0$ while the second requires $\Lambda=0$. This indicates that static solutions  require non-zero curvature; this is shown below for the Bianchi IX case. 

Of interest for a numerical study of eqns. (\ref{heffk}-\ref{pikdot}) is the dynamics of Kasner exponents (\ref{Kexp}); these are constants of the motions for the vacuum equations with $\Lambda=0$, but not otherwise. In particular, it is of interest to compare their dynamics in the classical and  classical-quantum cases as the singularity is approached and as the universe expands and approaches isotropy.

Numerical solutions of eqns. (\ref{heffk}-\ref{pikdot}) are shown in Fig. \ref{fig3} and \ref{fig4}  for the initial data $a_1 = 15.0$, $a_2 = 14.0$, $a_3 = 4.0$, $\pi_1 = -0.2$, $\pi_2 = -0.1$, with $m=0.001$ and $\Lambda =0$, and initial scalar states $|0\rangle$ and $(\sqrt{3}/2)|0\rangle + (1/2)|1\rangle$ respectively; $\pi_3$ is determined by solving ${\cal H}_{\rm eff}=0$, and the initial data for the classical scalar field case is again determined by eqn. (\ref{cdata}).  The first frame shows the evolving Kasner exponents  for the CQ case, including one of the sum rules; isotropization occurs as the universe expands, as for the classical case, but there are differences in both the volume and density at large volume. The energy density $\rho/\rho_0$ for the CQ case approaches a constant at large volume whereas the energy density of the C case continues to decline. Both the volume and energy density are larger for the CQ case at late times. 

Fig. \ref{fig5} shows the difference in the evolution of the  C and CQ exponents $p^{C}_k- p^{CQ}_k, \ k=1,2,3$ and the difference between their sum of squares for the data used for Fig. \ref{fig4}; while these differences are small, their cumulative affect on volume and energy density is evident at late time in the volume and scalar energy density mentioned above.

The differences between Figs. \ref{fig3} and \ref{fig4} arise solely from the different initial states of the scalar field. In both cases the expectation value of the scalar density $\rho =\langle\hat{{\cal H}}_\phi \rangle/v$ differs from the classical $\rho$; the latter shows some oscillation and decays to zero, whereas the former does not and appears to decay at a much slower rate (up to the integration times shown).  This is potentially significant in that it suggests the emergence of a ``cosmological constant" or equivalent ``dark energy" apparently caused by excitation of higher states of the scalar field at late times together with isotropization. The other main difference is in the volume, which at late times is larger than for the classical equations for both initial scalar states. 

Lastly, the other main difference between Figs. \ref{fig3} and \ref{fig4} is the excitation of higher energy levels of the scalar field as the universe expands. This is the homogeneous analog of particle creation; the initial state $|0\rangle$ shows excitation of only the even levels (since the hamiltonian is quadratic in creation/annihilation operators) whereas the the initial state   $\sqrt{3}/2)|0\rangle + (1/2)|1\rangle$ shows excitation of all levels. The Hilbert space truncation level used was $n=50$ for both figures.

\subsection{CQ Bianchi Universes}

The difference between Kasner and the other Bianchi models is the addition of spatial curvature. This is best studied in Misner's parametrization \cite{misner1969mixmaster} where the universe's volume is separated from its anisotropy variables; the curvature arises as a function of the latter through a gravitation potential term in the Hamiltonian constraint \cite{Ryan:1975jw}. Misner's parametrization is a transformation of the metric variables $(a_1,a_2,a_3)$ to $(\Omega, \beta_-,\beta_+)$:
\bea 
\Omega &=&-\frac{1}{3}\ln\left(a_1a_2a_3\right)\\
\beta_+&=&\frac{1}{3}\ln\left(\frac{a_1a_2}{a_3^2}\right),\\ \label{betaplus}
  \beta_-&=&\frac{1}{\sqrt{3}}\ln\left(\frac{a_1}{a_2}\right).
\eea 
 The conjugate momenta are related by 
\bea
p_\Omega&=&-(ap_a+bp_b+cp_c),\\
p_+&=&ap_a+bp_b-2cp_c,\\
p_-&=&\sqrt{3}(ap_a-bp_b).
\eea
In terms of these variables, the CQ effective Hamiltonian constraint is
\bea
H_{\rm eff} &=&\frac{e^{3\Omega}}{24}\left(p_+^2+p_-^2-p_\Omega^2\right)+e^{-3\Omega}\Lambda\nn\\
&& +e^{-\Omega}V(\beta_+,\,\beta_-)+\langle \hat{H}_\phi \rangle_{\psi}\nn\\
&=&0,
\label{Bianchihamconstraint}
\eea
where
\bea 
\langle \hat{H}_\phi \rangle_\psi= \frac{e^{3\Omega}}{2} \langle \hat{p}_\phi^2  \rangle_\psi+\frac{1}{2}m^2e^{-3\Omega} \langle \hat \phi^2  \rangle_\psi.
\eea
The spatial curvature term is $e^{-\Omega}V(\beta_-,\beta_+)$ which is determined by the Bianchi type summarized in the table below. The canonical equations of motion are
\bea
\label{omdot}
\dot \Omega &=& \{\Omega,\, H\}=-\frac{e^{3\Omega}}{12}\,p_\Omega\\
\label{betadot}
\dot \beta_\pm &=&\{\beta_\pm,\, H\}=\frac{e^{3\Omega}}{12}\,p_\pm\\
\label{pdot}
\dot p_\pm &=&\{p_\pm,\, H\}=-e^{-\Omega}\frac{\partial }{\partial \beta_\pm}V(\beta_+,\, \beta_-)\\
\label{pomdot}
\dot p_\Omega &=&\{p_\Omega,\, H\}=-3(H_K
-e^{-3\Omega}\Lambda)\nn\\
&&+ e^{-\Omega}V(\beta_+,\,\beta_-)-\frac{\partial }{\partial \Omega}\langle \psi | \hat H_\phi| \psi \rangle. 
\eea
where the last term expands as
\bea
\frac{\partial}{\partial \Omega} \langle \hat{H}_\phi \rangle_\psi &=& 
\frac{1}{\dot{\Omega}}\left(\dot{\langle \psi}| \hat{H}_\phi |\psi\rangle + \langle \psi| \hat{H}_\phi \dot{|\psi\rangle}\right) 
  + \langle \frac{\partial \hat{H}_\phi}{\partial \Omega}  \rangle_\psi \nn\\
&=& \frac{3}{2} \left(e^{3\Omega}\langle\hat{p}_\phi^2\rangle_\psi  -m^2 e^{-3\Omega}\langle\hat{\phi}^2\rangle_\psi\right).
\eea
 \vskip0.5cm

\begin{tabular}{|c|c|}
\hline
Bianchi Type & $V(\beta_+,\,\beta_-)$ \\
\hline
I & 0 \\
\hline
II & $e^{-8\beta_+}$ \\
\hline
III & $4e^{-(2\beta_+-2\sqrt{3}\beta_-)}$ \\
\hline
IV & $e^{4\beta_+}(12+e^{4\sqrt{3}\beta_-})$ \\
\hline
V & $12e^{4\beta_+}$ \\
\hline
VIII & 
$\begin{aligned}
 & e^{-8\beta_+}+2e^{4\beta_+}\left[\cosh(4\sqrt{3}\beta_-)-1\right]\\
&+4e^{-2\beta_+}\cosh(2\sqrt{3}\beta_-)
\end{aligned} $
\\
\hline
IX & 
 $\begin{aligned}
&e^{-8\beta_+}+2e^{4\beta_+}\left[\cosh(4\sqrt{3}\beta_-)-1\right]\\
&-4e^{-2\beta_+}\cosh(2\sqrt{3}\beta_-) 
\end{aligned}$\\
\hline
\end{tabular}
\vskip0.5cm

We now check if there are static solutions for any of the Bianchi universes  by setting metric functions constant and all momenta zero: 
\bea
\Omega &=& \Omega_0, \quad \beta_+=\bar \beta_+, \quad \beta_-=\bar \beta_-,\nn\\
p_\Omega&=&p_+=p_-=0,\label{classicstatickasner}
\eea
in (\ref{Bianchihamconstraint}) and (\ref{omdot}-\ref{pomdot}) with $|\psi\rangle=\sum_{k=1}^N c_k |k\rangle$, a linear combination of eigenstates of the quantum scalar field hamiltonian. This gives the conditions 
\bea
 e^{-3\Omega_0}\Lambda+e^{-\Omega_0}V(\bar\beta_+, \,\bar \beta_-)\nn\\
  +\ m \sum_{k=0}^N |c_k|^2\left(k+ 1/2\right)&=&0,\\
 V,_{\beta_\pm}&=&0,\\
 3e^{-\Omega_0}\Lambda+e^{-\Omega_0}V(\bar\beta_+, \,\bar \beta_-)&=&0;
 \label{BianchiStatic}
\eea
all other equations vanish identically, as does the last term in (\ref{pomdot}). 

From this it is evident that static solutions of the CQ equations require an extremum of the Bianchi potential $V$; with this, the last equation fixes the value of $\Lambda$, and the first fixes the expectation value of scalar Hamiltonian. Extrema of the potential occurs only for the Bianchi IX potential, and for this case, the extremum is a minimum. Hence, the static solution is stable.  

Lastly, we note that requiring a static solution of the  corresponding classical equations involves setting $\phi=$ constant and $p_\phi=\dot{p}_\phi=0$. For the massive scalar field, this imposes $\phi=0$; hence there is no classical static solution for the same Bianchi potential. 

 This result on the existence of static solutions of the Bianchi IX model shows that there can be significant differences between CQ and classical models. An intuitive check of linear stability suggests that the directions corresponding to anisotropies are stable since the Bianchi IX has a minimum at $\beta_\pm=0$; however, like the FRW CQ equations, there will be an unstable mode in the volume-state variables. This discussion is a first step toward a more complete numerical investigation of the Bianchi IX CQ equations (to be pursued elsewhere).

\section{Summary and Discussion} 

We studied a ``semiclassical" hybrid  geometrodynamics defined by first-order hamiltonian equations for classical geometry coupled to time-dependent Schrodinger equation for the state of a quantum scalar field. This system incorporates full backreaction of the quantum matter on classical geometry in a self-consistent manner.

 We next applied the idea to study the dynamics of homogeneous isotropic and anisotropic cosmology with a truncation of the scalar field Hilbert space. We showed that there are static solutions for both cases with certain conditions, and that these solutions are linearly unstable.   

 That static spacetimes can arise in the semiclassical Einstein equation is curious from the physical point of view. It suggests that the backreaction of the (global) quantum matter state on a classical spacetime, through the effective Hamiltonian constraint, can create a static classical-quantum bound state. From a mathematical perspective, the fact that the equations are non-linear in the state opens up the possibility of unusual solutions; such have been noted in the semiclassical spin-oscillator model \cite{Husain:2022kaz}. 

For the isotropic case we studied numerically the full evolution of initial data that is a perturbation of the semiclassical static universe parameters, for both the classical-quantum and classical equations. This revealed some similarities and differences; the latter arise due to the oscillations of the classical scalar field at early times, and in the late time decay of the energy densities. Of particular interest is that the free data for the  classical-quantum  system is the initial scale factor and scalar quantum state. This raises the question of whether the issue of fine tuning of scalar field initial data of the classical system can be avoided in the classical-quantum equations; only an initial quantum state is required rather than initial values of the scalar field and its momentum.  
 
That the past of an inflationary classical-quantum universe has  a static 3-sphere attractor in the past may be compared with the purely classical ``emergent universe" model proposed in \cite{Ellis:2002we}, where the scalar field has an arbitrary scalar potential (unlike only the mass term here). Our analysis provides a similar model for the origin of inflation starting from a classical-quantum model with full backreaction.

It is useful to contrast our FLRW results with other works on the instability of emergent universes \cite{Mithani:2012ii,Mithani:2014jva,QuantInstability}: the first of these works argues that in a minisuperspace quantization without matter, an emergent universe would collapse; the second rules out oscillating solutions in loop quantum cosmology as the origin of an emergent universe; the third, unlike our case, considers the evolution of a scalar field wave packet on a fixed background (i.e. no backreaction).  Including non-perturbative backreaction (as we do here) results in a static universe that has an unstable mode; it is this mode that leads to inflation, while the other mode decays exponentially back to the semiclassical static universe.  

For the aniotropic case we show that unstable static universes also arise for the Bianchi IX model; this is because one of the conditions for static solutions is the requirement of critical points of the curvature potential. 

Thus, while we have shown that the semiclassical static universe derived here can inflate or stay stable, the question of how it might arise  from quantum gravity remains, as does the validity of the semiclassical Einstein equation as a transitional theory between quantum gravity and quantum theory on curved spacetime; a recent discussion of this issue appears in \cite{Husain:2022kaz}, and a linear alternative is  proposed in \cite{Oppenheim:2018igd}.

The numerical evolution of the CQ equations for the isotropic and Kasner cases shown in Figs. \ref{fig1}-\ref{fig4} show  significant differences from classical evolution, especially concerning volume and energy density---for the CQ equations  both quantities are larger than that for the classical equations. In particular, the scalar energy density energy tends to a constant, a fact following from the expression 
\bea
\langle \hat{\rho}\rangle_\psi = \frac{\langle \hat{p}_\phi^2\rangle_\psi}{2v^6} +\frac{m^2}{2} \langle\hat{\phi}^2\rangle_\psi;
\eea
the second term dominates at late times and does not vanish, unlike for the classical scalar field. This suggests a possible contribution to dark energy if the CQ equations apply at sufficiently late times.  

An important question is the domain of validity of the CQ approximation. In the absence of a quantum theory of gravity, this is hard to assess. However there is indication from  models such as the oscillator-spin \cite{Husain:2022kaz} and the oscillator-oscillator systems \cite{Husain:2023jaq} that the CQ approximation is valid only for weak coupling and only for sufficiently small times. In the quantum gravity of cosmological models, which are solvable in several cases, a similar comparative investigation of the full quantum and  CQ regimes is possible. (In field theory, the case of scalar quantum electrodynamics  provides another testing ground for the regime of validity of CQ approximation: this has been partially investigated in \cite{Kiefer:1991xy} in the case of an evolving vacuum state.) 

Beyond cosmological models, the Hamiltonian approach to semiclassical theory applied here opens up the possibility of studying the classical gravity-quantum scalar field theory in spherical symmetry---a setting of particular interest for gravitational collapse generalizing the classical case \cite{Choptuik:1992jv}, and for backreaction of Hawking radiation. A possible intuitive lesson from our study of the cosmological CQ system is that late time behaviour can differ significantly with full backreaction due to the the expectation values of the $\hat{\phi}^2$.

Another possibility for further study concerns the CQ
equations themselves. Here these were postulated. Can they be derived from the full quantum theory of the system with reasonably justified approximations? The case of coupled oscillators studied recently \cite{Husain:2023jaq} provides an approach to consider for constrained Hamiltonian systems like gravity.

\smallskip


 \begin{acknowledgments}
 We thank George Ellis, Irfan Javed, Roy Maartens, Sanjeev Seahra, and Edward Wilson-Ewing for helpful comments and references. This work was supported by the Natural Science and Engineering Research Council of Canada. VH thanks the Perimeter Institute where this work was completed; Research at Perimeter Institute is supported in part by the Government of Canada through the Department of Innovation, Science and Economic Development Canada and by the Province of Ontario through the
25 Ministry of Colleges and Universities.
\end{acknowledgments}

\bibliography{references.bib}  

\end{document}